# Optical properties of silicon rich silicon nitride ($Si_xN_yH_z$) from first principles.

**Shu Xia Tao**[a,b*], **Anne M.M.G. Theulings**[a,b], **Violeta Prodanović**[a,b], **John Smedley**[c], **Harry van der Graaf**[a,b]

[a] *National Institute for Subatomic Physics, Nikhef Science Park 105, 1098 XG Amsterdam, The Netherlands*
[b] *Delft University of Technology TNW, Mekelweg 15, 2629 JB Delft, The Netherlands*
[c] *Brookhaven National Laboratory, Upton, New York 11973, USA*

*E-mail sxtao@nikhef.nl

ABSTRACT:

The real and imaginary parts of the complex refractive index of $Si_xN_yH_z$ have been calculated using density functional perturbation theory. Optical spectra for reflectivity, adsorption coefficient, energy-loss function (ELF), and refractive index, are obtained. The results for $Si_3N_4$ are in agreement with the available theoretical and experimental results. To understand the electron energy loss mechanism in Si rich silicon nitride, the influence of the Si doping rate, of the positions of the dopants, and of H in and on the surface on the ELF have been investigated. It has been found that all defects, such as dangling bonds in the bulk and surfaces, increase the intensity of the ELF in the low energy range (below 10 eV). H in the bulk and on the surface has a healing effect, which can reduce the intensity of the loss peaks by saturating the dangling bonds. Electronic structure analysis has confirmed the origin of the changes in the ELF. It revealed that the changes in ELF is not only affected by the composition but also by the microstructures of the materials. The results can be used to tailor the optical properties, in this case the ELF of Si rich $Si_3N_4$, which is essential for secondary electron emission application.

INTRODUCTION

Silicon nitrides are commonly used for high-temperature and high-endurance applications due to their excellent mechanical strength and wear resistance [1]. Recent progress of micro electro-mechanical systems (MEMS) technology in the semiconductor industry has led to significant advancements in the development of silicon-nitride based microelectronic technology, including as an insulator [2,3], as a diffusion barrier to prevent diffusion of impurities, as a masking material for KOH etching [4], and for masking to prevent oxidation as used in the LOCOS process [5].

Meanwhile, the advantages of smaller, less costly systems have provided opportunities for researchers to develop versatile MEMS devices on the microscale. One of these MEMS based devices is a novel photon detector with a goal of ps temporal resolution proposed by H. van der Graaf [6] at NIKHEF, Amsterdam. The photon detector (called Tipsy) being developed has the potential to revolutionize photon detection with its superb temporal resolution. This device relies on a stacked set of curved miniature thin (in order of 10 nm) silicon nitride films for generating secondary electrons, replacing the traditional gain dynodes in a photo multiplier tube. The main challenge of realizing Tipsy is sufficient secondary electron yield (SEY) at relative low primary electron energy. For operation, a yield of 4 secondary electrons must be achieved at 500 eV.

MEMS based silicon nitride is expected to be a promising secondary electron emission material for Tipsy for four reasons: 1) Silicon nitride has a wide band gap (5.1 eV [7]), where high SEY can be expected. 2) There is one experimental report of SEY of 2.9 at reflective mode at primary energy of 350 eV for LPCVD silicon nitride thin films [8]. 3) Mechanically stable and ultrathin (in order of 10 nm) silicon nitride films can be realized through MEMS technology [9]. For such thin systems, the SEY from above (reflection mode) and from below (transmission mode) the film is expected to be equally high. 4) Various surface treatments may induce negative electron affinity on silicon nitride surfaces, such as H termination or alkali metal oxide termination, according to our previous prediction by DFT calculations [10, 11].

Despite the potential advantages of silicon nitride as a secondary electron emission material, there are few literature reports on their secondary electron emission properties. Furthermore, there are few systematic studies on the electron energy loss mechanism in silicon nitride. Within the research group of H. van der Graaf [6], recent efforts have been made in this matter by using a Monte Carlo code based on the Geant4 platform of CERN [12], which is developed by Bosch and Kieft [13]. This code is one of few with a good description of low-energy (<50 eV) interactions of electrons with matter. For the modelling of inelastic scattering (electron scattering with energy loss) events, dielectric function theory is applied to calculate the inelastic mean free path. Dielectric function theory allows derivation of differential cross-sections for electron–atom interaction from optical data.

The advantage of using the dielectric function theory is that optical data are readily available for a wide range of materials, and in broad energy ranges. However, despite there were several literature reports for $Si_3N_4$, the data have generally been limited to the evaluation of the index of refraction in the visible region and the infrared measurements. The focus of these experiments have been used mainly for structural and chemical evaluation purposes [14]. Other studies also indicate the presence of chemically bound hydrogen in the film. Thus, the stoichiometry of an arbitrary film can be more accurately indicated by $Si_xN_yH_z$. Thus the refractive index and other optical properties will be a function of x, y and z. To study secondary electron emission property of silicon rich silicon nitride, a complete set of optical data (ELF) in a wide energy range is necessary.

The purpose of this paper is to use density functional perturbation theory (DFPT) to perform a systematic investigation of the optical properties of Si rich silicon nitride in a wide energy range (from infrared to ultraviolet). The real and imaginary parts of the complex refractive index have been calculated. Optical spectra for reflectivity, adsorption coefficient, energy-loss function (ELF), and refractive index are obtained, and the results for pure β-$Si_3N_4$ are validated against available theoretical and experimental results. The materials have been studied in the form of $Si_xN_yH_z$ (modified based on the β-$Si_3N_4$), so that the effect of the Si doping rate, of the position of the Si dopants, and of the H in the bulk and on the surfaces can be taken into account. Furthermore, density of states (DOS) analysis has revealed that the change in the ELF has a direct link to the electronic structure of the materials. This work opens a window into the application in the fields of secondary electron emission by having an insight into the dielectric and optical properties of Si rich silicon nitride.

**COMPUTATIONAL METHODS AND STRUCTURAL MODELS**

Geometry optimization and electronic structure calculations were performed before the optical spectrum calculations. All calculations were performed using DFT as implemented in the Vienna Ab-Initio Simulation Package (VASP) [15, 16]. The Kohn–Sham equations were solved using a basis of Projector Augmented Wave functions with a plane-wave energy cut-off of 400 eV [17, 18]. The electron-exchange correlation energy is described by using the functional of Perdew, Burke, and Ernzerhof (PBE) based within the generalized gradient approximation (GGA) [19]. Detailed computational setups, lattice parameters of bulk β-$Si_3N_4$, and their validation by comparing with literatures [20-24] are described in our previous study [10].

To model Si rich silicon nitride in the form of $Si_xN_yH_x$, modifications of the used unit cell have been made based on that of β-$Si_3N_4$. For example, $Si_{13}N_{15}$ has been modelled by replacing one N atom by one Si atom in a unit cell with 28 atoms; $Si_7N_7$ has been modelled by replacing one N by one Si atom in the unit cell with 14 atoms. For $Si_7N_7$, two different distributions of the Si dopants have been included: Si distributed homogenously or forming a cluster (four N atoms next to each other are replaced by four Si atoms) in the unit cell. To study the influence of the H in the bulk Si rich silicon nitride on the ELF, three types of H in the bulk in $Si_7N_7$, namely, H forming bonding with only Si, only N, both Si and H, respectively. Geometric and electronic structures of clean and H terminated (10 $\bar{1}$ 0) and (11 $\bar{2}$ 0) β-$Si_3N_4$ surfaces have been documented in our previous study [10] and details will not be shown here again.

The optical calculations of this work is based on density functional perturbation theory with the independent particle approximation (IPA). The implementation in VASP is the calculation of the response function in the framework of the PAW method. Details are given in references [25, 26]. In crystals of high symmetry, the nomenclature of the isotropic medium can still be applied along high symmetry directions. So in the hexagonal unit cells β-$Si_3N_4$, the components of the dielectric tensor parallel to *a*, *b* and *c*-axes are inspected.

VASP calculates the frequency dependent dielectric matrix after the electronic ground state has been determined. The imaginary part is determined by a summation over empty states using the equation:

$$\varepsilon_{\alpha\beta}^{(2)}(\omega) = \frac{4\pi^2 e^2}{\Omega} \lim_{q \to 0} \frac{1}{q^2} \sum_{c,v,k} 2w_k \delta(\varepsilon_{ck} - \varepsilon_{vk} - \omega) \times \langle u_{ck+e_\alpha q} | u_{vk} \rangle \langle u_{ck+e_\beta q} | u_{vk} \rangle^* . \quad (1)$$

Where the indices c and v refer to conduction and valence band states respectively, and $\mu_{ck}$ is the cell periodic part of the wave functions at the k-point k. The real part of the dielectric tensor is obtained by the usual Kramers-Kronig transformation.

$$\varepsilon_{\alpha\beta}^{(1)}(\omega) = 1 + \frac{2}{\pi} P \int_0^\infty \frac{\varepsilon_{\alpha\beta}^{(2)}(\omega')\omega'}{\omega'^2 - \omega^2 + i\eta} d\omega' . \quad (2)$$

The main optical spectra, such as the reflectivity $R(\omega)$, adsorption coefficient $I(\omega)$, energy-loss spectrum $L(\omega)$, and refractive index $n(\omega)$, all can be obtained from the dynamical dielectric response functions $\varepsilon(\omega)$. The explicit expressions are given by

$$R(\omega) = \left| \frac{\sqrt{\varepsilon(\omega)} - 1}{\sqrt{\varepsilon(\omega)} + 1} \right|^2, \quad (3)$$

$$I(\omega) = (\sqrt{2})\omega \left[ \sqrt{\varepsilon_1(\omega)^2 + \varepsilon_2(\omega)^2} - \varepsilon_1(\omega) \right]^{1/2}, \quad (4)$$

$$L(\omega) = \varepsilon_2(\omega) / \left[ \varepsilon_1(\omega)^2 + \varepsilon_2(\omega)^2 \right], \quad (5)$$

and

$$n(\omega) = (1/\sqrt{2}) \left[ \sqrt{\varepsilon_1(\omega)^2 + \varepsilon_2(\omega)^2} + \varepsilon_1(\omega) \right]^{1/2}. \quad (6)$$

**RESULTS AND DISCUSSION**

**Validation of β-Si₃N₄**

First of all, the static dielectric tensor of β-Si₃N₄ has been determined using IPA. They are 4.21 (parallel to *a* and *b*) and 4.29 (parallel to *c*). This is in good agreement with previous theoretical values of 4.19 and 4.26 from Cai et al [27] and of 4.24 and 4.32 from Watts [28]. Figure 1 shows the imaginary and real part of the frequency dependent dielectric tensor as a function of photon energy in the energy range of 0 - 60 eV with components parallel to (*a, b*) and *c*-axes. The imaginary part of the frequency dependent dielectric tensor represents the absorption spectrum of a material. β-Si₃N₄ is transparent in the energy range from 0.0 to about 5.0 eV and is featured by main absorption peaks between 8.5 eV and 10.5 eV, which continuously slopes down with increasing photon energies. It has slightly anisotropic characteristics parallel to the *c*-axis due to the hexagonal unit cells. Main optical spectra, i. e. the reflectivity, adsorption coefficient, energy-loss function (ELF), and refractive index can be calculated from the frequency dependent dielectric tensor shown in Figure 1. Due to the lack of experimental data for other spectrums, only the refractive index and ELF have been plotted and compared with experimental references in Figures 2 and 3, respectively.

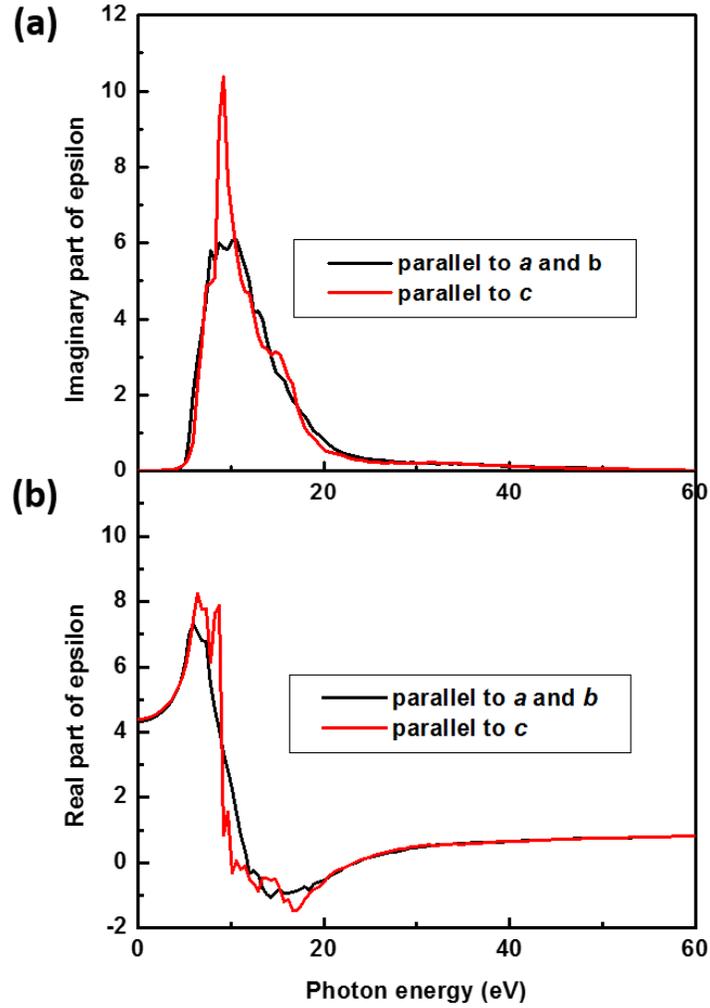

Figure 1. The frequency dependent dielectric function ε (ω) = $\varepsilon_1$ (ω) + i $\varepsilon_2$ (ω) of β-$Si_3N_4$ as a function of the photon energy. (a) and (b) represent our calculated imaginary and real parts of dielectric function ε (ω), respectively. The black and red lines are polarization parallel to (*a*, *b)* and *c*-axes, respectively.

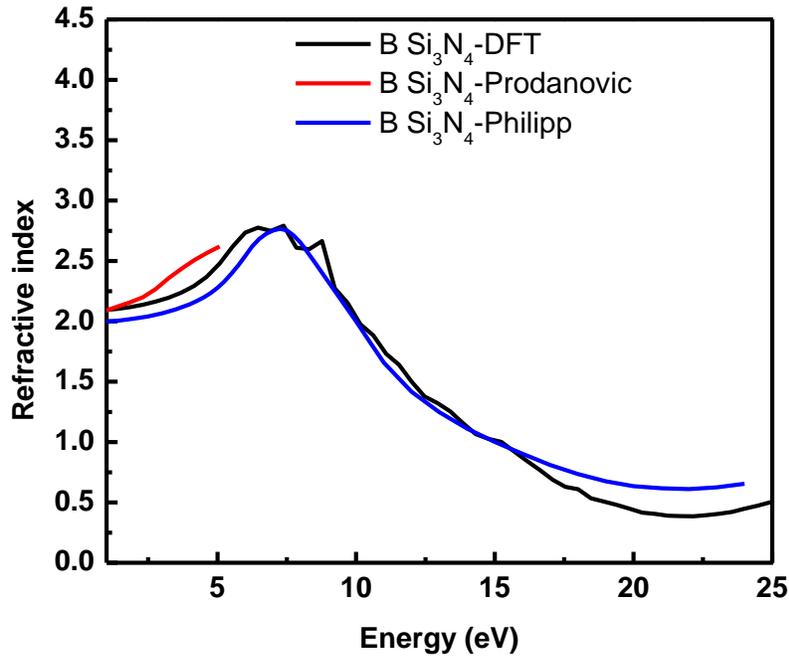

Figure 2. A comparison of the refractive index as a function of photon energy. The black curve represents the DFT results for β-Si3N4, and the red and blue lines are optical measurements for amorphous Si3N4. The blue line is reproduced from Philipp [29]. The blue line was obtained by spectroscopic ellipsometry by Prodanovic [30] from the same research group.

For the refractive index (see Fig. 2), the calculated curve is nicely positioned in the middle of the two experimental curves within the available energy range. The discrepancy between experimental and theoretical data are smaller than the discrepancy between the experimental data (due to the differences in the preparations of the samples, the measurement techniques and the derivation of the data). For the ELF (see Fig. 3), the agreement is also very good except two main differences between the DFT and experimental plots: i) there is a kink at the peak just above 20 eV in the experimental curve, which is very likely caused by combining different sets of optical data from different sources due to lack of experimental data in a wide energy range from the same work [14]; ii) The peaks below 0.3 eV seen in experimental curve is absent in the one obtained from DFT. The latter can be explained by the fact that the optical spectra are directly calculated from inter-band transitions, where the phonon modes at very low energies are not included. As shown by Monte Carlo studies carried out by the same authors, the secondary electron emission properties of β-Si3N4 is barely affected by the inclusion of this part of the ELF. [31]

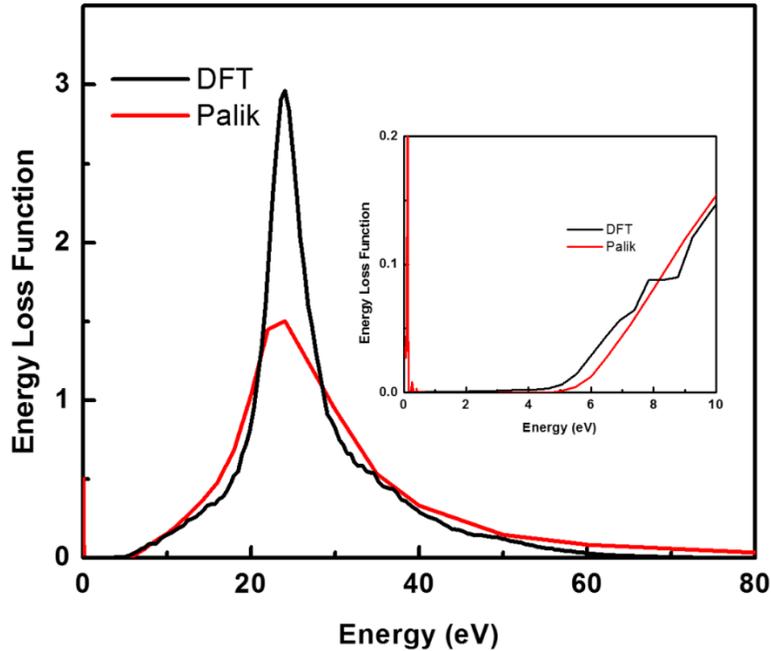

Figure 3. A comparison of the energy loss spectrum as a function of photon energy. The black curve represents the DFT results for β-$Si_3N_4$, and the red curve represents the curve obtained from the combination of a few sets of optical data from handbook of Palik[14].

**$Si_xN_y$ – the influence of the Si doping rate and positions**

Figure 4 illustrates the calculated ELF of Si rich silicon nitride in comparison with those of β-$Si_3N_4$ and Si. The studied forms of Si rich silicon nitride are $Si_{13}N_{15}$ (one N atom is replaced by one Si in a unit cell with 28 atoms), $Si_7N_7$ (one N atom is replaced by one Si atom in the unit cell with 14 atoms), and $Si_7N_7$ cluster (the four N atoms are replaced by four Si atoms forming a Si cluster inside the unit cell with 56 atoms). Compared to that of β-$Si_3N_4$, the ELF of the Si doped silicon nitride shows three main features:
1) The general shape and the position of the main peaks are similar, where the ELF increases from low energy and peaks at about 23 to 27 eV and slopes down at higher energy.
2) The main peaks of the Si rich silicon nitride have shifted slightly to lower energy (at about 1-2 eV). Therefore, the magnitude of the ELF at low energies side of the slope is larger than that of β-$Si_3N_4$.
3) Extra energy loss peaks appear below 5 eV. The increase in the magnitude of the ELF and number of peaks are larger for highly Si doped silicon nitride ($Si_7N_7$) than for lower Si doped silicon nitride ($Si_{13}N_{15}$).

For the influence of the Si dopant position, one should compare the ELF of the $Si_7N_7$ and $Si_7N_7$ cluster. Compared with the ELF of $Si_7N_7$, that of $Si_7N_7$ cluster exhibits more Si character which can be summarized as follows:

1) The position of the main peak of $Si_7N_7$ cluster is significantly sharper and is in a position at about 5 eV lower, and which is closer to that of Si.
2) The positions of the adsorption peaks below 10 eV can be seen as a combination of those of Si, $Si_3N_4$, and $Si_7N_7$, but the peas are in general smaller and broader (below 1 eV and at about 7 eV).

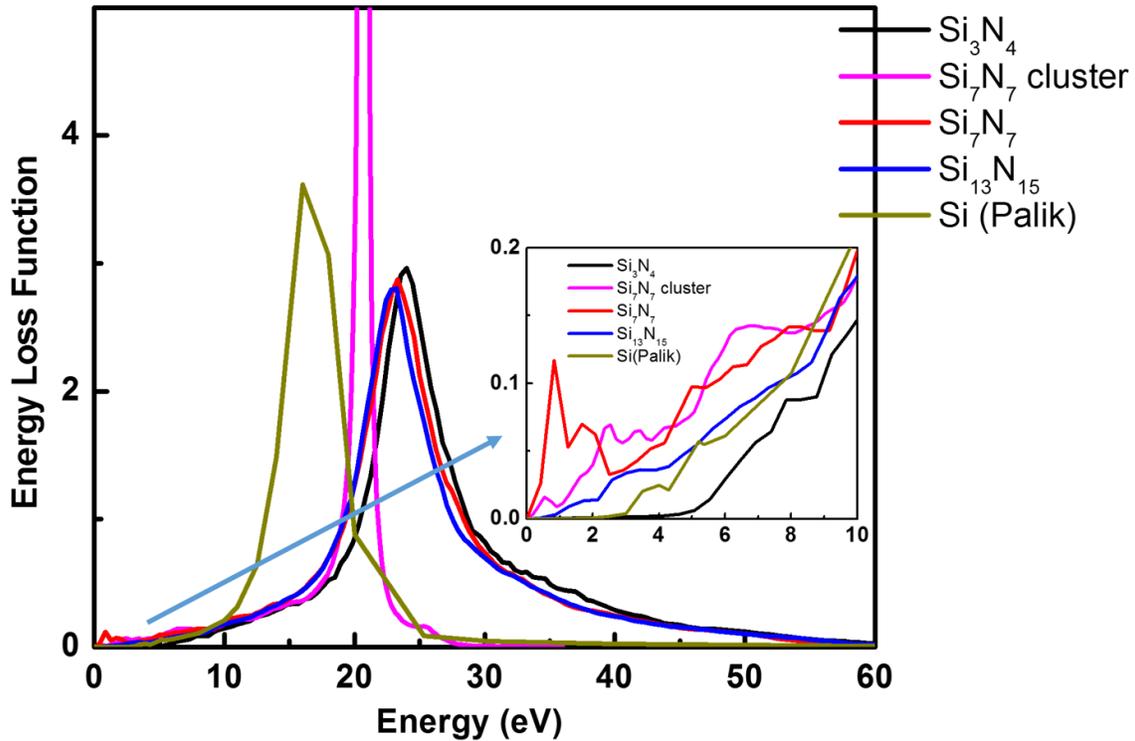

Figure 4. Energy loss function of Si rich silicon nitride $Si_xN_y$ with comparison with $\beta$-$Si_3N_4$ and Si obtained from experimental reference [14].

Since the optical spectra are directly calculated from interband transitions, a comparison of the electronic structures should be useful to understand the changes in the ELF. Shown from the density of states (DOS) in Fig 5, the excess Si in Si rich silicon nitride $Si_xN_y$ has introduced i) extra electronic states in the band gap just above the Fermi level which can be assigned to the unsaturated Si bonds, ii) broader Si nonbonding feature in the range of 5-10 eV at conduction band. This observation is similar to a recent finding of Hintzsche [32] for amorphous $Si_xN_yH_z$. For the $Si_7N_7$ cluster, the intensity of the unsaturated Si feature is much smaller and the overall DOS plot is shifted to lower energy.

To summarize Figures 4 and 5 together, the extra energy loss peaks at $Si_7N_7$ below 5 eV are due to the introduction of the electronic states at the band gap originating from the unsaturated Si atom. For the $Si_7N_7$ cluster case, the ELF can be seen as a combination of Si and $Si_3N_4$ and Si rich $Si_3N_4$.

Based on this calculated result, one can conclude that the optical properties of the Si rich silicon nitride not only are determined by the composition of the materials, but also are determined by the microstructure in an atomic level.

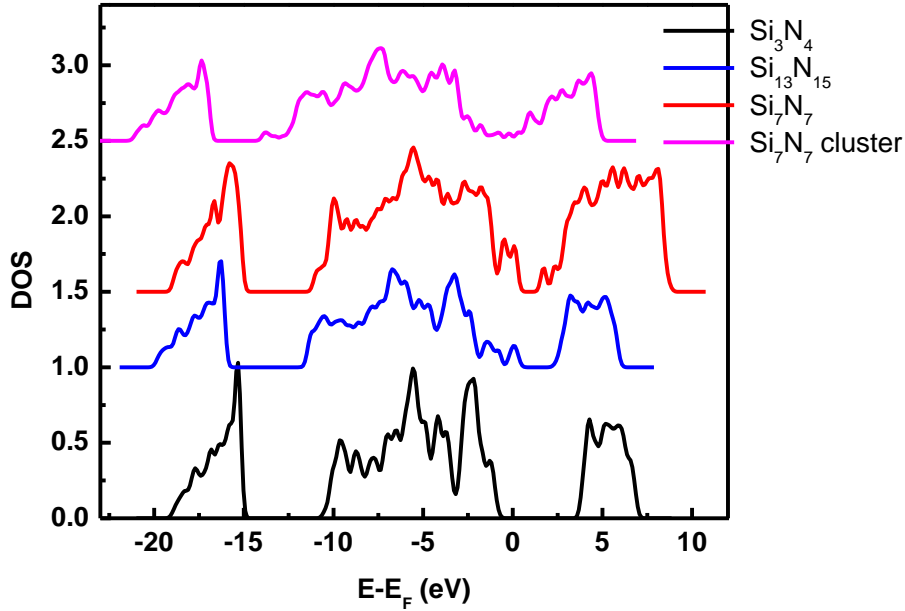

Figure 5. Density of states (DOS) of Si rich silicon nitride ($Si_xN_y$), compared to β-$Si_3N_4$.

**$Si_xN_yH_z$ –H in bulk Si rich silicon nitride**

Figures 6 and 7 summarize the ELF and DOS of $Si_xN_yH_z$ in the form of $Si_7N_7H_{1-2}$ in comparison with $Si_7N_7$. Three possibilities have been studied. They are one H forming bond with the dangling bond of the access Si, one H with N, and two H with both Si and N, respectively. The main peaks of the ELF of $Si_7N_7H_{1-2}$ are located at almost the same energies as that of $Si_7N_7$. The main differences can be found in the energy range below 5 eV. The sharp peak located at about 1 eV for $Si_7N_7$ is now absent for all the three, where the H has saturated directly or indirectly the defects in the Si rich silicon nitride. This function of H in curing defects is the largest when bonded with Si, the smallest when bonded with N, and logically intermediate for a combination of the two.

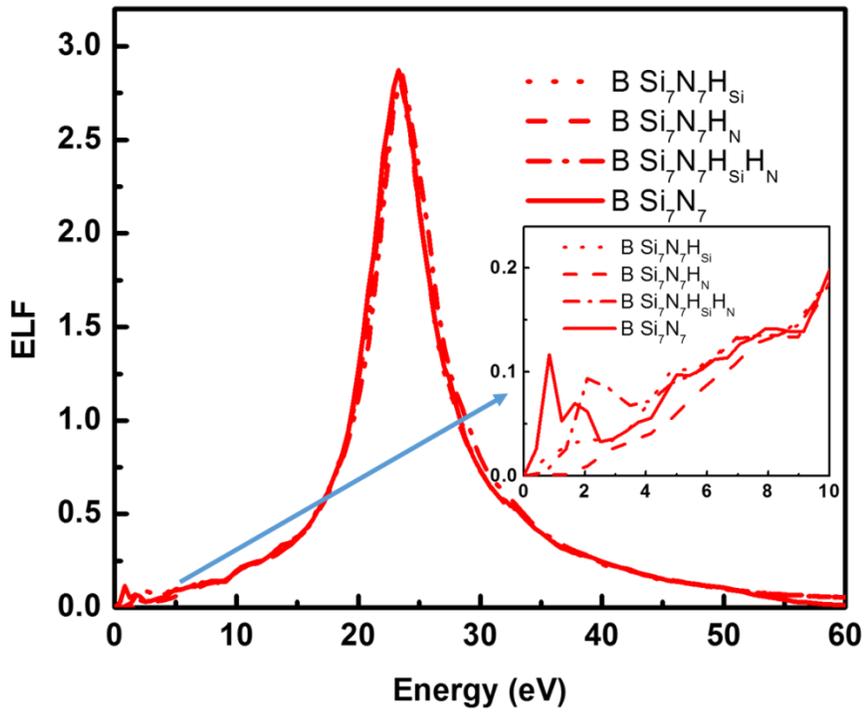

Figure 6. Energy loss function of H in Si rich silicon nitride $Si_7N_7H_{1-2}$ in comparison with $Si_7N_7$. $Si_7N_7H_{Si}$, $Si_7N_7H_N$, and $Si_7N_7H_NH_{Si}$ represent H forming bonds with unsaturated Si atoms, N atoms, and the combination of the two, respectively.

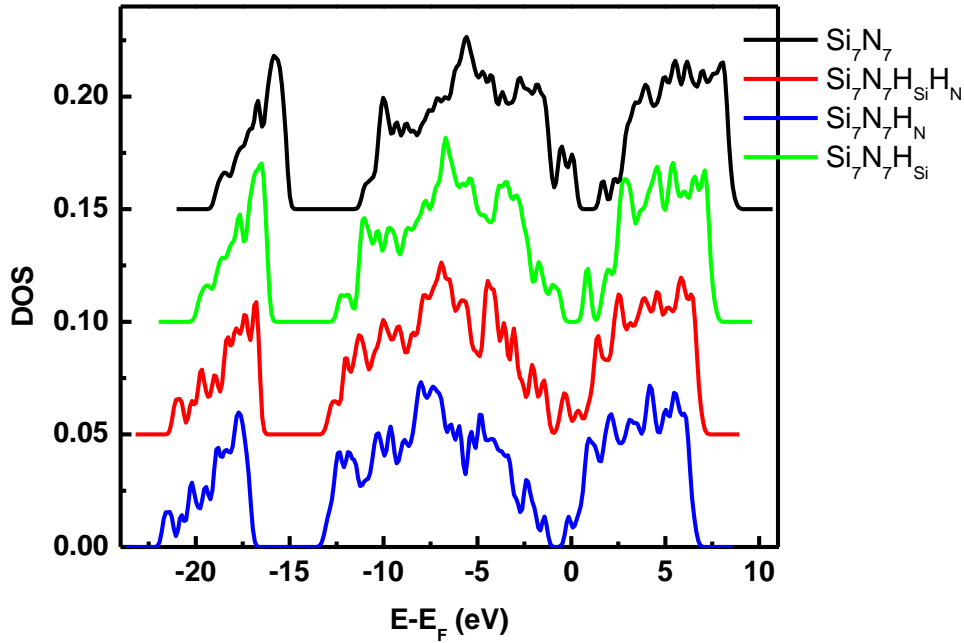

Figure 7. Density of States (DOS) of Si rich silicon nitride ($Si_7N_7H_{1-2}$) in comparison with $Si_7N_7$.

**$Si_xN_yH_z$ –H termination on the surfaces of $Si_3N_4$**

**Surface vs bulk**

The effect of surface defects has been studied by comparing bulk β-$Si_3N_4$ with two surfaces: (10$\bar{1}$0) and (11$\bar{2}$0). Compared with bulk β-$Si_3N_4$, the main peak of the ELFs (see Fig. 8) of both surfaces have shifted energies which are about 3 to 5 eV lower. Similarly with defects in the bulk, the defects introduced by surfaces also lead to adsorption peaks at about 4 eV. However the peaks in the system with surface defects are much lower and broader that the ones observed in system with bulk defects. This is because in both β-$Si_3N_4$ with surfaces and Si rich silicon nitride, defects (unsaturated bonds) occur and these introduces extra electronic states as seen in the DOS (see Fig. 9). However, the defects on the surface are more often self-cured in the sense that the atoms on the surfaces tend to relocate to such positions to saturate the dangling bonds. In the bulk there is limited volume to allow the atoms to do so.

**Clean surface vs H termination**

Again, similar to H in the bulk, H termination on the surfaces reduces the defect effect by saturating the dangling Si and N bonds. This can be seen in the slightly smoother ELF and DOS plots, in particular in the energy range of 0-5 eV.

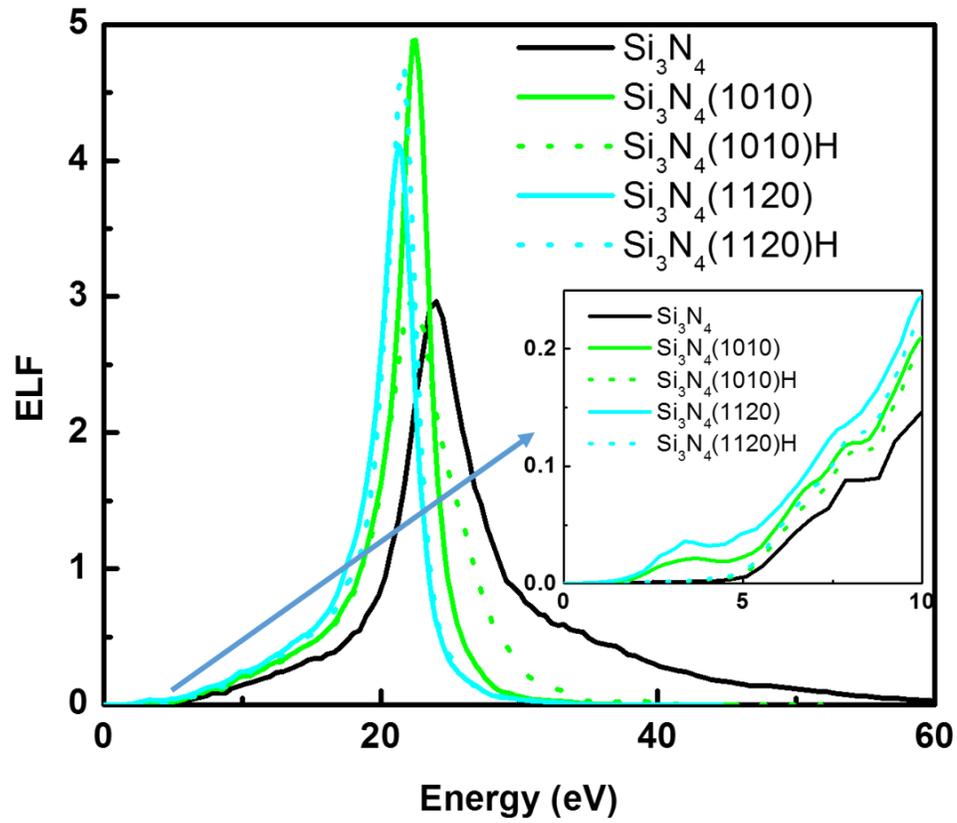

Figure 8. Energy loss function of surfaces with and without H termination in comparison with bulk β-$Si_3N_4$.

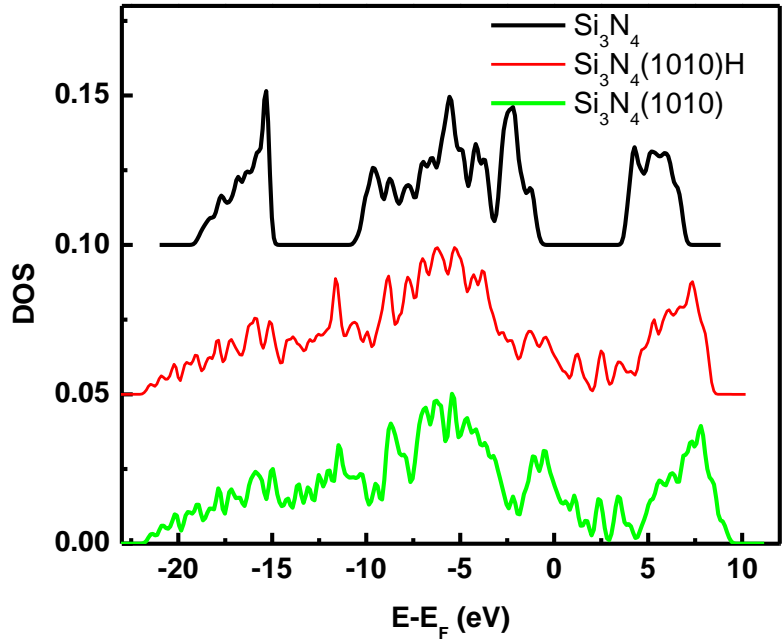

Figure 9. Density of States (DOS) of surfaces with and without H termination in comparison with bulk β-$Si_3N_4$.

Monte Carlo simulations using the calculated ELF have been performed to evaluate the influences of the material parameters (such as doping rate, surface defects, H terminations, and electron affinity) on the secondary electron yield (SEY) of thin Si rich silicon nitride films. The results indicate that the SEY is extremely sensitive to the ELF in low energies, especially below 10 eV. The highest SEY is obtained using the ELF of the bulk β-$Si_3N_4$. This is mainly due to the large band gap, where the secondary electrons are more likely to escape to the surface without losing energy from interacting with impurities. As expected, the Si doping in silicon rich silicon nitride and surfaces both cause decrease in the SEY. Whereas H adsorption in the bulk and H termination on the surface both have reduced the effect caused by the defects. Details of the electron transport mechanism and SEY results will be published in a separate paper [31]. By combining with Monte Carlo simulations, the significance of this work is to provide a guideline for tuning the material parameters to achieve favorable secondary electron emission properties.

The other optical spectra (reflectivity, adsorption coefficient and refractive index) of all the studied systems are shown in Figures 10, 11 and 12 for the sake of completeness. Details will be not analyzed here since the focus of this work is the ELF.

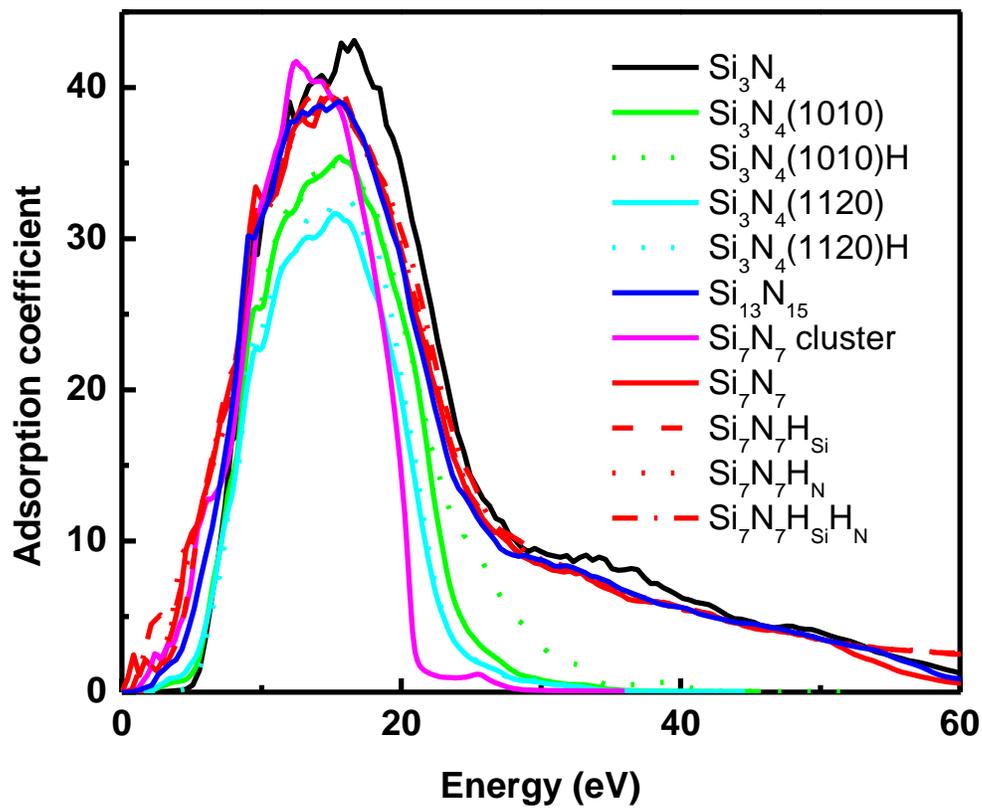

Figure 10. Adsorption coefficient of $Si_xN_yH_z$.

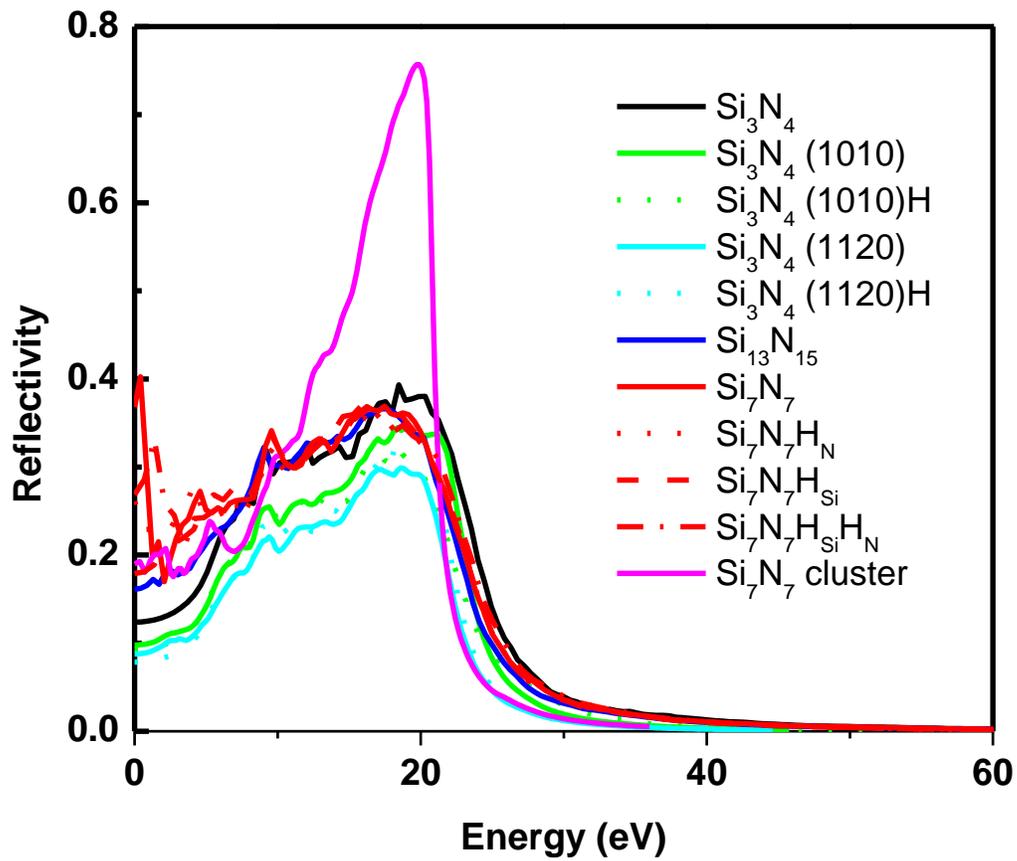

Figure 11. Reflectivity of $Si_xN_yH_z$.

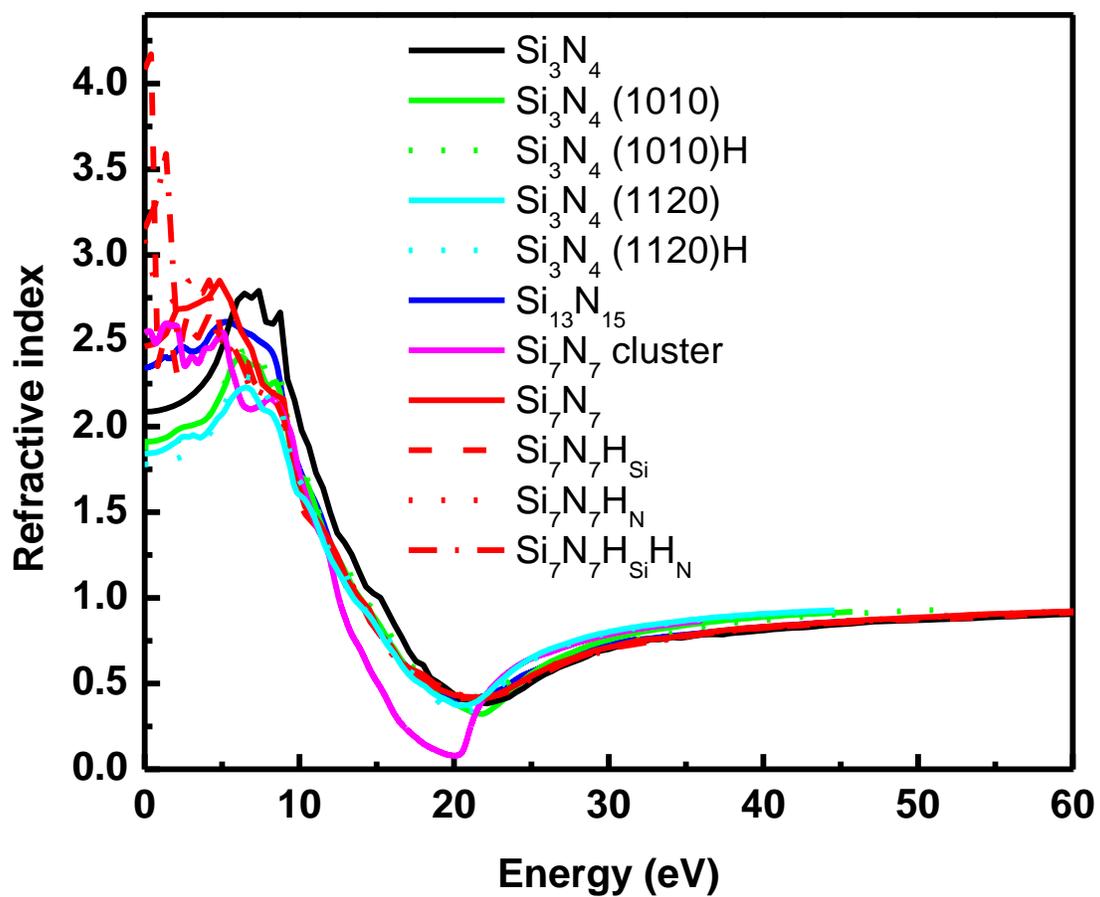

Figure 12. Refractive index of $Si_xN_yH_z$.

**Summary:**

   The optical properties of $Si_xN_yH_z$ have been studied by DFPT, with a focus on the energy loss spectrum. Defects in the bulk (due to the excess Si) and on the surface, and the influence of hydrogen in the bulk, hydrogen termination on the surfaces have been investigated. Extra energy loss peaks have been found in the energy range lower than 10 eV in all the studied systems compared with $\beta$-$Si_3N_4$. The origins of these energy loss peaks have been identified in their corresponding electronic structures. It is concluded that the changes in ELF is not only affected by the composition but also by the microstructures of the materials. The excess Si in the bulk in Si rich silicon nitride leads to the strong increase both in the magnitude and the number of adsorption peaks in ELF, whereas those caused by surface defects are much milder. H in the bulk of Si rich silicon nitride and H termination on surfaces have healing effect by saturating the dangling bonds and therefore reduce the number and flatten the sharpness of the adsorption peaks. The results can be used as reference data for tuning the optical properties via controlling the composition and microstructures for secondary electron emission applications.

**ACKNOWLEDGEMENTS:** This work is funded by the European Research Council (320764) (ERC-Advanced 2013 MEMBrane).

[30] Optical properties of thin silicon nitride layers are obtained by spectroscopic ellipsometry (J. A. Woollam Ellipsometer M2000UI). This is a widely used technique based on polarization analysis of an incident light beam and a beam being reflected by investigated sample. A special mathematical model is built for the measured sample based on laws of reflection and refraction in thin films. Measurement of two quantities, Ψ and Δ, is performed in the center of 4-inch wafer with a silicon nitride film deposited on a silicon substrate. Spectra is acquired for 5 different angles in range 50° - 70° with a step of 5°. Finally, fitting is performed to find a set of parameters which provides a match between generated and measured data. Thicknesses of layers are verified also by SEM inspection of cross section of samples.

[31] A. Theulings, S.X. Tao, H. van der Graaf, Monte Carlo simulation of low energy photon-electron interaction and transportation in silicon rich silicon nitride, in preparation.

[32] L. E. Hintzsche, C. M. Fang, M. Marsman, G. Jordan, M. W. P. E. Lamers, A. W. Weeber, and G. Kresse, Defects and defect healing in amorphous $Si_3N_{4-x}H_y$: An ab initio density functional theory study, PRB 88, (2013) 155204.